\documentclass[conference]{IEEEtran}
\IEEEoverridecommandlockouts
\usepackage{cite}
\usepackage{amsmath,amssymb,amsfonts}
\usepackage{algorithmic}
\usepackage{graphicx}
\usepackage{textcomp}
\usepackage{xcolor}
\usepackage{booktabs} 

\usepackage[capitalise]{cleveref}

\def\BibTeX{{\rm B\kern-.05em{\sc i\kern-.025em b}\kern-.08em
    T\kern-.1667em\lower.7ex\hbox{E}\kern-.125emX}}
\begin{document}

\title{Sensor Data Augmentation from Skeleton Pose Sequences for Improving Human Activity Recognition
}

\author{\IEEEauthorblockN{
            Parham Zolfaghari\IEEEauthorrefmark{1}\IEEEauthorrefmark{2}\thanks{P. Zolfaghari and V. Fortes Rey - These authors contributed equally to this work.},
            Vitor Fortes Rey\IEEEauthorrefmark{2}\IEEEauthorrefmark{3},
            Lala Ray\IEEEauthorrefmark{2}\IEEEauthorrefmark{3},
            Hyun Kim\IEEEauthorrefmark{4},
            Sungho Suh\IEEEauthorrefmark{2}\IEEEauthorrefmark{3}\thanks{Corresponding author: sungho.suh@dfki.de},
            Paul Lukowicz\IEEEauthorrefmark{2}\IEEEauthorrefmark{3}}
        \IEEEauthorblockA{\IEEEauthorrefmark{1}Department of Computer Science, Saarland University, Saarbrücken, Germany}
        \IEEEauthorblockA{\IEEEauthorrefmark{2}German Research Center for Artificial Intelligence (DFKI), Kaiserslautern, Germany}
        \IEEEauthorblockA{\IEEEauthorrefmark{3}Department of Computer Science, RPTU Kaiserslautern-Landau, Germany}        
        \IEEEauthorblockA{\IEEEauthorrefmark{4}Department of Electrical and Information Engineering, Seoul National University of Science and Technology, Korea}}


\maketitle


\begin{abstract}

The proliferation of deep learning has significantly advanced various fields, yet Human Activity Recognition (HAR) has not fully capitalized on these developments, primarily due to the scarcity of labeled datasets. Despite the integration of advanced Inertial Measurement Units (IMUs) in ubiquitous wearable devices like smartwatches and fitness trackers, which offer self-labeled activity data from users, the volume of labeled data remains insufficient compared to domains where deep learning has achieved remarkable success. Addressing this gap, in this paper, we propose a novel approach to improve wearable sensor-based HAR by introducing a pose-to-sensor network model that generates sensor data directly from 3D skeleton pose sequences. our method simultaneously trains the pose-to-sensor network and a human activity classifier, optimizing both data reconstruction and activity recognition. Our contributions include the integration of simultaneous training, direct pose-to-sensor generation, and a comprehensive evaluation on the MM-Fit dataset. Experimental results demonstrate the superiority of our framework with significant performance improvements over baseline methods.
\end{abstract}



\section{Introduction}
Human activity recognition (HAR) has emerged as a cornerstone technology in a myriad of applications, ranging from personal fitness to healthcare and industrial automation. The ubiquity of smart wearable devices, such as watches and fitness trackers, has made continuous monitoring of physical activities not only possible but also prevalent. These devices, equipped with advanced sensors, provide a rich source of data that, when analyzed, can offer personalized health and fitness recommendations, guiding users toward their wellness goals.

Beyond fitness, the implications of precise HAR are profound, extending into elderly care for fall detection \cite{yoshida2022data}, patient monitoring in healthcare \cite{sangeethalakshmi2023patient}, and even worker safety in manufacturing line \cite{suh2023worker}. Accurate activity recognition facilitates the development of intelligent systems that are responsive to human needs, enhancing safety, productivity, and overall quality of life.

Despite its potential, HAR faces significant challenges, chiefly due to the limited availability of labeled datasets. In contrast to computer vision tasks, a significant bottleneck is the scarcity of annotated sensor data required for training accurate and robust recognition models. Manually labeling extensive datasets for diverse human activities is time-consuming, labor-intensive, and expensive. Additionally, deploying sensors on multiple body parts, though essential for capturing comprehensive motion data, exacerbates the complexities and costs associated with data collection and annotation.

To address these challenges, researchers have explored alternative avenues. Data augmentation stands as a powerful technique to combat the dearth of labeled data, enhancing the diversity and volume of training datasets, and thereby improving the performance of machine learning models. In scenarios where data is scarce or expensive to obtain, augmentation has proven to be effective, particularly in image and speech recognition tasks. In particular, recently, several sensor data generation methods from video sequences have emerged in the HAR community. Existing works \cite{kwon2020imutube, fortes2021translating, santhalingam2023synthetic} estimate 2D or 3D joint positions from videos and infer joint orientations to compute inertial measurement unit (IMU) data. While these methods improved the spatial accuracy of generated sensor data and enhanced the performance of wearable sensor-based HAR, they still struggle with capturing the intricacies of sensor characteristics and may not fully exploit the potential for cross-modality transfer. 

In this paper, we propose a novel approach to improve wearable sensor-based HAR performance. We introduce a pose-to-sensor network model that generates IMU sensor data directly from 3D skeleton pose sequences. Unlike existing approaches, the proposed method trains the pose-to-sensor network and the HAR classifier at the same time. The pose-to-sensor network is trained to minimize the reconstruction loss between the ground truth of sensor data and generated sensor data and the classification loss of the human activity classifier with real and generated data. This concurrent training process allows the pose-to-sensor network to generate synthetic sensor data that not only closely resembles real sensor data but also enhances the performance of the activity classifier. By minimizing both the reconstruction loss between ground truth and generated sensor data and the classification loss of the HAR classifier, our method leverages the synergistic relationship between these components to optimize overall performance. 
To evaluate the proposed method, we evaluate the proposed method on the well-known open-access benchmark datasets, MM-Fit \cite{stromback2020mm} and UTD-MHAD \cite{chen2015utd}, which provide multimodal data including the skeleton, IMU, and label data, enabling the synthesis of realistic and diverse training samples. Experimental results demonstrate that the proposed framework provides better performance improvement in terms of accuracy and macro F1 score compared to existing methods, including IMUTube \cite{kwon2020imutube}, Chen et al. \cite{chen2015utd}, and Memmesheimer et al. \cite{memmesheimer2020gimme}, and baseline which trains the pose-to-sensor network model without considering the classifier. 


The main contributions of our proposed method are summarized as follows.

\begin{itemize}
    \item We introduce a novel approach that integrates the training of the pose-to-sensor network with a human activity classifier, promoting a synergistic optimization that improves both data reconstruction and activity recognition.
    \item By directly generating sensor data from 3D skeleton pose sequences, the proposed method addresses the limitations of existing methods that map pose joints to specific sensor positions, potentially capturing more subtle sensor characteristics for HAR.
    \item We conduct comprehensive evaluations on the MM-Fit \cite{stromback2020mm} and UTD-MHAD \cite{chen2015utd} datasets to demonstrate the superiority of the proposed framework. Comparative analyses showcase significant performance improvements over existing methods and baseline methods that solely focus on pose-to-sensor network training without considering the classifier.
\end{itemize}

The remainder of this paper is organized as follows: \cref{sec:relatedwork} reviews related work improving HAR performance by generating IMU sensor data augmentation. \cref{sec:method} describes the methodology, including the data synthesis process and the end-to-end training pipeline. \cref{sec:experiments} presents the experimental setup, results, and comparative analysis. Finally, \cref{sec:conclusion} concludes the paper with a summary of our findings and a discussion of potential future work.


\section{Related Works}
\label{sec:relatedwork}

Recently, generative models have been utilized to improve the performance of wearable sensor-based HAR by augmenting sensor data \cite{li2020activitygan, hu2023bsdgan, zuo2023unsupervised}. In particular, several notable methods have emerged in the literature to facilitate the generation of virtual IMU data from video sequences. Generative methods, such as \cite{rey2019let, fortes2021translating}, have employed machine learning techniques to derive IMU data from videos directly. In addition, trajectory-based approaches, as demonstrated by \cite{kwon2020imutube, xiao2021deep}, extracted 2D joint positions from videos and estimated 3D pose joint positions from 2D pose sequences. Subsequently, they estimated joint orientations using forward kinematics. The resulting orientations enable the transformation of 3D joint positions into frame-of-reference of the IMU, facilitating the computation of acceleration and angular velocity. 

In response to challenges associated with labeled data acquisition, the concept of virtual IMU data generation has gained traction. Cross-modality transfer approaches, such as those proposed by \cite{rey2019let, fortes2021translating, kwon2020imutube}, extract virtual IMU data from 2D RGB videos of human activities. This not only addresses limitations in labeled wearable data collection but also contributes to the construction of personalized HAR systems for individual user needs \cite{xia2022virtual}. Virtual IMU data generation improves the accuracy and robustness of HAR models, promoting broader adoption across diverse domains.

Among the innovative systems, IMUTube \cite{kwon2020imutube} stands out as a comprehensive solution for extracting virtual IMU data from 2D RGB videos. Operating as a processing pipeline, IMUTube integrates computer vision, graphics, and machine learning models to convert large-scale video datasets into virtual IMU data suitable for training sensor-based HAR systems. Its adaptive video selection, 3D human motion tracking, and virtual IMU data extraction and calibration components collectively contribute to generating high-quality virtual IMU data. Notably, the system's versatility has been demonstrated in improving model performance through the integration of real and virtual IMU data. 
However, existing approaches primarily focus on generating real-like IMU data from pose sequences or video data, neglecting the potential to improve the performance of the activity classifier. These methods lack a simultaneous training approach that optimizes the pose-to-sensor network model to minimize both reconstruction loss and classification loss, limiting their ability to fully leverage the available data for enhanced activity recognition.

In this landscape, the proposed method in this paper introduces a novel approach to sensor data generation from 3D skeleton pose sequences, focusing on the simultaneous training of a pose-to-sensor network model and a human activity classifier. Unlike existing methods that primarily identify corresponding sensor data from specific sensor positions on human body parts, the proposed approach aims to optimize the pose-to-sensor network model by minimizing both reconstruction loss and classification loss. 
By addressing these limitations and innovatively leveraging 3D skeleton pose sequences, our proposed method offers a promising solution for enhancing the performance and robustness of wearable sensor-based HAR systems. Subsequent sections introduce the details of this methodology and present experimental results, demonstrating its superior performance compared to baseline approaches that neglect classifier considerations.


\begin{figure*}[!t]
  \centering
  \includegraphics[width=\textwidth]{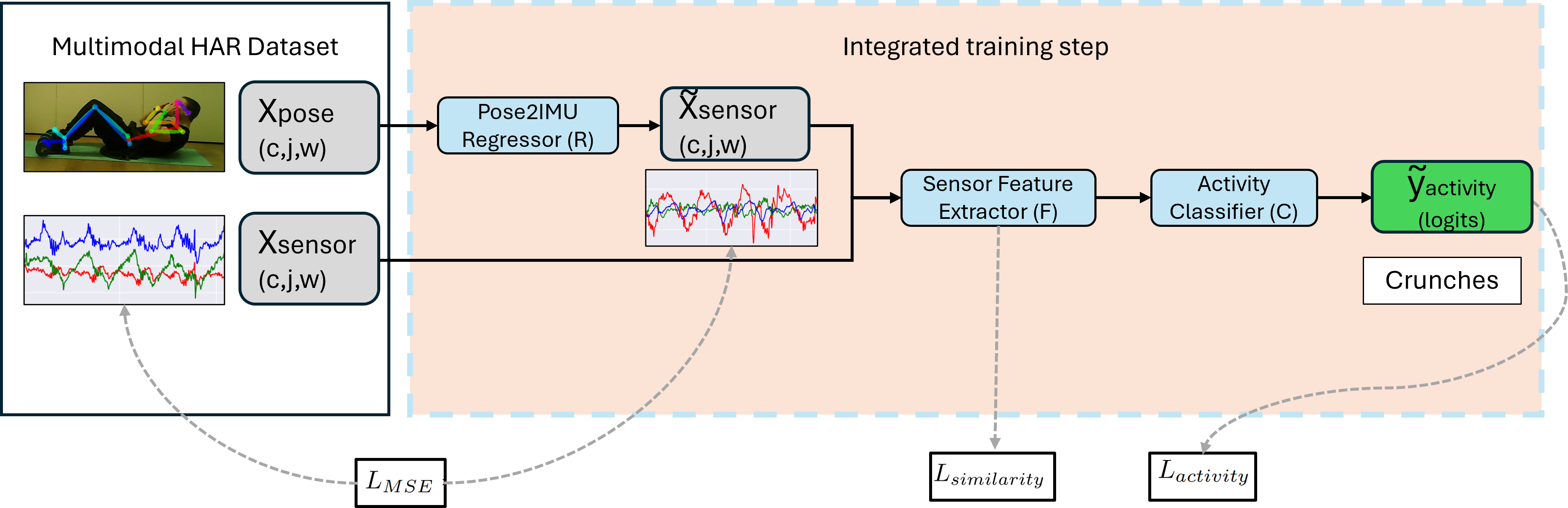}
  \caption{Schematic Overview of the Proposed Method. This diagram illustrates the training architecture wherein the feature extraction and classification modules are concurrently trained with both real and synthetic IMU accelerometer data. Integral to the pipeline is a regression model that generates synthetic data and facilitates the enhanced training of the feature extraction and classification modules. The overall training process is governed by a compound weighted sum loss, optimizing the synergy between the modules for improved performance}
  \label{fig:scenario2}
\end{figure*}

\section{Method}
\label{sec:method}
This section describes our method to enhance HAR which focuses on the efficient use of available real and simulated sensor data.
While most other methods consider sensor data simulation and training a HAR classifier as different, independent steps, ours performs both in an end-to-end fashion.

Initially, we detail the preprocessing steps applied to prepare the data for training, ensuring its suitability for the subsequent learning process. We then introduce the architecture of our end-to-end pipeline, which concurrently leverages synthetic and real IMU data to enhance the training of the HAR model.

\subsection{Pre-processing}
The pre-processing stage is critical for synchronizing and standardizing our data sources. Thus, first, we match the sampling rate between the pose data and the sensor one. As poses are obtained from the video, which can have frame rates of 30 or below, while IMUs can reach higher sampling rates such as 100 Hz, we bring both modalities to a fixed 100 samples per second using linear interpolation.
Subsequently, we standardize the accelerometer data to a mean of zero and a standard deviation of one per channel, using statistics obtained in the training set. Simultaneously, we normalize the skeleton data to ensure uniformity across the dataset.
Normalization of the skeleton data is performed using the method proposed by Rey et al. \cite{fortes2021translating}. We calculate the Euclidean distance between the neck and mid-hip joints at a single time $t$, denoted as $dist_t$. To mitigate the effect of outliers, this distance is computed over a window $w$ of three seconds, and the scale at time $t$ is determined by the median of distances within this window:
\begin{equation}
    scale_t = median(dist_{t-w/2}, ..., dist_{t+w/2})
\end{equation}
and we define a function to scale any value $v_t$
\begin{equation}
    scale(v_t, scale_t) = -1.0 + \frac{v_t}{scale_t} * 2.0
\end{equation}
We compute new values for each joint setting the mid-hip joint as a reference
\begin{equation}
    NewJoint^t = scale(joint^t - midhip^t, scale_t)
\end{equation}
Since the data structure of the UTD-MHAD dataset is different from that of MMFit by having separate files and already synchronized for each class we simply applied linear interpolation to the 30 Hz skeleton data to match the 50 Hz inertial data of UTD-MHAD.
Afterward, the standard normalization method listed above is followed to prepare the final data. 

\subsection{End-to-end pipeline}
The cornerstone of our proposed method is an end-to-end pipeline designed to enhance the training efficacy of the feature extractor module, thereby yielding more accurate activity classification. Central to this pipeline is a regression model that employs temporal convolutional network (TCN) blocks \cite{bai2018empirical} as its core. This model is tailored to process the 3D positions of the number of joints outputting synthesized IMU accelerometer sensor data of a joint.

The components of our method are illustrated in \cref{fig:scenario2}. Our approach comprises various interconnected networks, all trained concurrently. The first component, a regression model denoted as $R$, receives pre-processed pose sequences as input and produces simulated sensor data. Formally, applying $R$ to the pre-processed pose data $x_{pose}$ generates our simulated sensor data $\tilde{x}_{sensor}$. To guide the model to generate realistic sensor data, we incorporate the mean squared error (MSE) term between real and simulated sensor data into the overall loss function:
\begin{equation}
    L_{MSE} = \lVert x_{sensor} - \tilde{x}_{sensor} \rVert^{2}
\end{equation}

Inspired by \cite{fortes2021translating}, we customized the architecture of our regression model to better process 3D joint position data. The core of our regression model comprises five TCN blocks, each followed by a linear layer. Two notable adjustments are made from the original TCN block design presented in Rey et al. \cite{fortes2021translating}. First, we employ 2D convolution layers instead of 1D, a change necessitated by our use of 3D joint positions as input data. Second, we substitute the ReLU activation function with Leaky ReLU for enhanced performance, applying it after two 2D convolutions alongside a dropout layer. Consistent with the configuration in \cite{fortes2021translating}, the initial TCN block in our model does not incorporate dropout. \cref{tab:regression} shows the full architecture of the proposed model.
In case of UTD-MHAD dataset we modified the architecture by removing the TCN Block 5 and adding a view layer to before the FC layer to adopt to the data shape.

\begin{table}[!t]
\centering
\caption{Regression model architecture}
\label{tab:regression}
\begin{tabular}{|c|c|}
\hline
Layer & \begin{tabular}[c]{@{}c@{}}TCN Block\\ (in\_ch, out\_ch, kernel\_size, \\ dilation, dropout)\end{tabular} \\ \hline
TCN Block 1 & (3, 32, 3, 1, 0) \\ \hline
TCN Block 2 & (32, 32, 3, 2, 0.2) \\ \hline
TCN Block 3 & (32, 32, 3, 4, 0.2) \\ \hline
TCN Block 4 & (32, 32, 3, 1, 0.2) \\ \hline
TCN Block 5 & (16, 16, 1, 1, 0.1) \\ \hline
Fully Connected & \begin{tabular}[c]{@{}c@{}}Linear(out\_features=3*window)\end{tabular} \\ \hline
\end{tabular}
\end{table}

\begin{table}[!t]
\centering
\caption{Feature Extraction Model Architecture and Classification}
\label{tab:feature_extraction}
\begin{tabular}{|c|c|}
\hline
Layer & Configuration \\ \hline
1D Conv & \begin{tabular}[c]{@{}c@{}}Input channels: 3,\\ Output channels: 9,\\ Kernel size: 9,\\ Stride: 9//2\end{tabular} \\ \hline
Leaky ReLU & - \\ \hline
1D Batch Norm & f = 9 \\ \hline
Dropout & 0.2 \\ \hline
1D Conv & \begin{tabular}[c]{@{}c@{}}Input channels: 9,\\ Output channels: 9,\\ Kernel size: 9,\\ Stride: 9//2\end{tabular} \\ \hline
Leaky ReLU & - \\ \hline
1D Batch Norm & f = 9 \\ \hline
Dropout & 0.2 \\ \hline
1D Conv & \begin{tabular}[c]{@{}c@{}}Input channels: 9,\\ Output channels: 9,\\ Kernel size: 9,\\ Stride: 9//2\end{tabular} \\ \hline
Leaky ReLU & - \\ \hline
1D Batch Norm & f = 9 \\ \hline
Dropout & 0.2 \\ \hline
1D Maxpool & Kernel size: 2, Stride: 2 \\ \hline
Fully Connected & f\_out = 100 \\ \hline
Fully Connected & f\_out = n classes \\ \hline
\end{tabular}
\end{table}

Both real and simulated sensor data are then processed by the same feature extraction $F$ module and then by our classifier $C$. We investigated promoting alignment between feature vectors derived from real and synthetic accelerometer data by including in our loss the cosine similarity between both, as expressed by:
\begin{equation}
    L_{similarity} = L_{CS}(F(x_{sensor}), F(\tilde{x}_{sensor}))  
\end{equation}
where $x_{sensor}$ and $\tilde{x}_{sensor}$ denote real sensor data and synthetic sensor data, respectively.

Simultaneously, for activity classification accuracy, we employ cross-entropy loss with both real and synthetic feature vectors, defined as:
\begin{equation}
\begin{split}
    L_{activity} = & -\sum_{l=1}^{N_{activity}} w_{i} y_l \log C ( F ( x_{sensor_l} ) ) \\
    & -\sum_{l=1}^{N_{activity}} w_{i} y_l \log C ( F ( \tilde{x}_{sensor_l} ) )
\end{split}
\end{equation}
Our feature extraction model consists of three 1D convolution layers with maxpooling after the first and second convolution layers, culminating in a linear layer that prepares the feature vector for the classification module. The classification is performed by a single-layer fully connected network. \cref{tab:feature_extraction} details the architecture of the feature extractor and classification models. In the case of the UTD-MHAD dataset we adopted the classification model from Chen et al. \cite{chen2015utd}.

Unlike conventional approaches, all networks of our proposed framework are trained concurrently. This concurrent training strategy, in which the regression model, feature extractor, and classifier are collectively optimized, plays a key role in smoothly integrating real and simulated data into a robust and effective system. The total loss function encapsulates the contributions of the reconstruction loss $L_{MSE}$ for sensor data generation, the activity classification loss $L_{activity}$ for activity classification, and the cosine similarity loss $L_{similarity}$ for feature alignment. 
\begin{equation}
    L_{final} = L_{MSE} + \alpha L_{activity} + \beta L_{similarity}
\end{equation}
where $\alpha$ and $\beta$ are weighting factors that balance the contribution of each loss component to the total loss.




\section{Experiments and Results}
\label{sec:experiments}
This section provides a comprehensive overview of the empirical evaluation conducted to assess the effectiveness of our proposed method. We commence by detailing the MM-Fit dataset, which serves as the foundational data source for our study. Following this, we outline the baseline method against which we benchmark our approach, establishing a context for comparative analysis.

This section culminates with a presentation of the results derived from our evaluation, illustrating the impact of our modifications and the overall performance of our proposed method in the context of Human Activity Recognition (HAR).


\subsection{Dataset}
The MM-Fit dataset forms the core of our study, tracking participants engaged in a variety of workout activities. It provides a comprehensive suite of data capturing 2D and 3D skeletal poses and Inertial Measurement Unit (IMU) sensor readings from the wrist and other body locations across 21 workout sessions. Each session is composed of three sets, with each set containing ten exercises and ten repetitions. During the rest intervals between sets, the sensors continue to record, capturing the natural rest behavior of participants. The dataset categorizes eleven different activities, including periods of 'no workout'.

The UTD-MHAD dataset offers a comprehensive collection of 27 diverse human actions that includes 20 upper body and 7 lower body motions, ranging from simple gestures like hand waves and clapping to complex movements like basketball shooting and jogging. A Kinect camera and wearable inertial sensor is used for data collection. The Kinect camera captures high-resolution color and depth images at a frame rate of approximately 30 frames per second, providing detailed visual data. Complementing this, the wearable inertial sensor, positioned either on the subject's wrist or thigh depending on the nature of the action, records precise motion data including acceleration, angular velocity, and magnetic strength at a sampling rate of 50 Hz.

For our research, we focus specifically on the 3D skeleton pose data, obtained from an RGB camera operating at 30 Hz, and the left wrist accelerometer data from a smartwatch with a 100 Hz sampling rate. As in \cite{fortes2021translating}, our regression model receives as input the three left-arm joints (wrist, elbow, and shoulder).

To ensure consistency and comparability with previous studies, we have adhered to the same training, validation, and test splits as employed in the original MM-Fit paper, testing on the fixed unseen users test set. We used 3-second sliding windows (300 samples) with a 0.2-second stride and assigned for each window the activity that happened the most during it as its label. This approach allows for direct comparison of our results with the established benchmarks in the field.



\subsection{Baselines}

This subsection outlines two baseline methods used for activity classification. The first baseline method utilizes only real accelerometer data from the left wrist. It involves a feature extraction model that identifies key features from the accelerometer data, which are then used by a classifier to predict the activity type. Both the feature extractor and the classifier follow the same architecture as our proposed method. This approach is akin to the one used in the MM-Fit \cite{stromback2020mm} study and serves as a comparison to demonstrate the enhancements our proposed method provides.

The second baseline is an approach where we first train a regression model to generate synthetic sensor data from 3D joint poses and then use the synthetic data provided by said model, together with real sensor data to train the activity classifier. By integrating synthetic data, this model aims to augment the training dataset and improve classification performance. However, unlike our proposed method, this process involves separate steps for data synthesis and classifier training.

Both baselines are critical for evaluating the effectiveness of our proposed method. By comparing these two approaches, we aim to showcase how including classification in the sensor generation procedure can improve both the quality of the generated sensor data as well as the overall classification performance for real data in terms of F1 score and accuracy, emphasizing our contributions to the field of Human Activity Recognition.


\begin{figure*}[ht!]
    \centering
    \includegraphics[width=\textwidth]{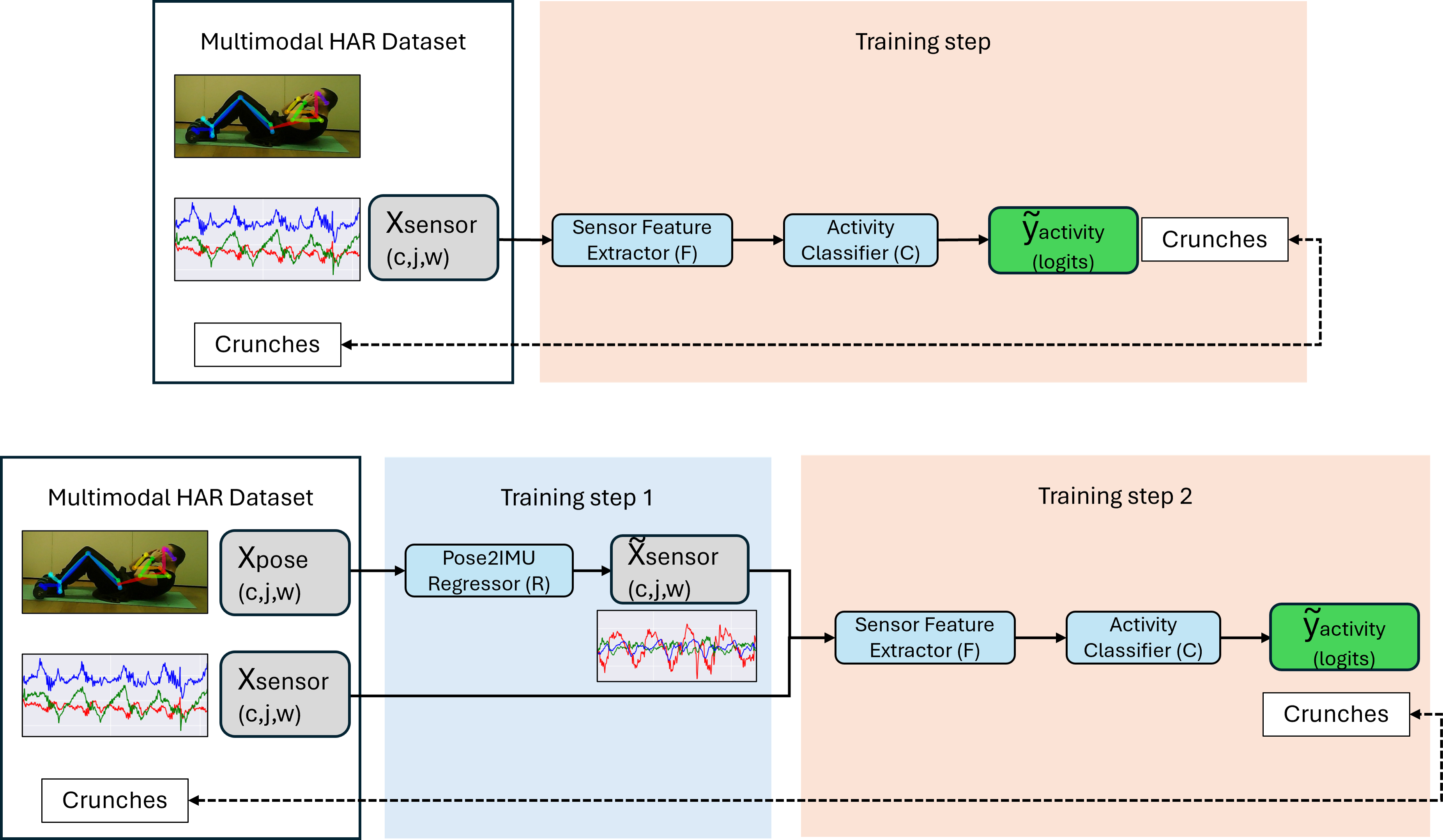}
    \caption{Overview of Baseline Approaches for Activity Classification. The first baseline utilizes only real sensor data (\( \mathbf{x_{\text{sensor}}} \)) for classifier training. The second baseline employs a two-step approach where a regression model (\( R \)) first predicts synthetic sensor data (\( \mathbf{\tilde{x}}_{\text{sensor}} \)) from 3D joint pose sequences (\(\mathbf{x}_{\text{pose}} \)), which is then combined with real data to train the classifier. Both methods converge at the Activity Classifier (\( C \)), which outputs the activity predictions (\( \mathbf{\tilde{y}}_{\text{activity}} \)).}
    \label{fig:my-diagram}
\end{figure*}

In our approach, we explore the efficacy of integrating both real and synthetic IMU accelerometer data to enhance the training of the feature extraction and classification modules. Initially, we developed a regression model trained independently to accurately predict IMU accelerometer data based on the arm's joint positions, specifically the wrist, elbow, and shoulder. Following the successful generation of synthetic accelerometer data, we merge it with the real sensor data. This combined dataset is then utilized to train the feature extraction and classification modules, aiming to leverage the diversity and comprehensiveness of the augmented data set for improved model performance.


\subsection{Implementation Details}
To ensure a consistent starting point for model training, we initialize the weights of the convolution and fully connected layers using Kaiming initialization. Recognizing the potential impact of initialization randomness, we conducted five runs of the experiment for MMFit and ten runs for UTD-MHAD each with a predefined seed to comprehensively explore the search space.
The Adam optimizer, with a learning rate of $10^{-3}$, was selected for training. To prevent over-fitting and ensure optimal generalization, we implemented an early stopping mechanism based on the F1 score on the validation set, with a patience parameter of 25 epochs for MMFit and 30 epochs for UTD-MHAD, respectively. The models were trained for a maximum of 100 epochs for MMFit and 200 epochs or until the early stopping criterion was met.

The introduction of early stopping, based on validation set performance, plays a critical role in our experimental design, allowing us to halt training when the model ceases to show improvement, thereby conserving computational resources and avoiding over-fitting.

The results from these experiments highlight the efficacy of our proposed modifications and the robustness of our end-to-end pipeline. The improvements in F1 score and accuracy, detailed further in this section, substantiate our hypothesis that integrating synthetic accelerometer data and optimizing the feature extraction process can significantly enhance HAR performance.


\begin{figure*}[!t]
  \centering
  \includegraphics[width=0.49\textwidth]{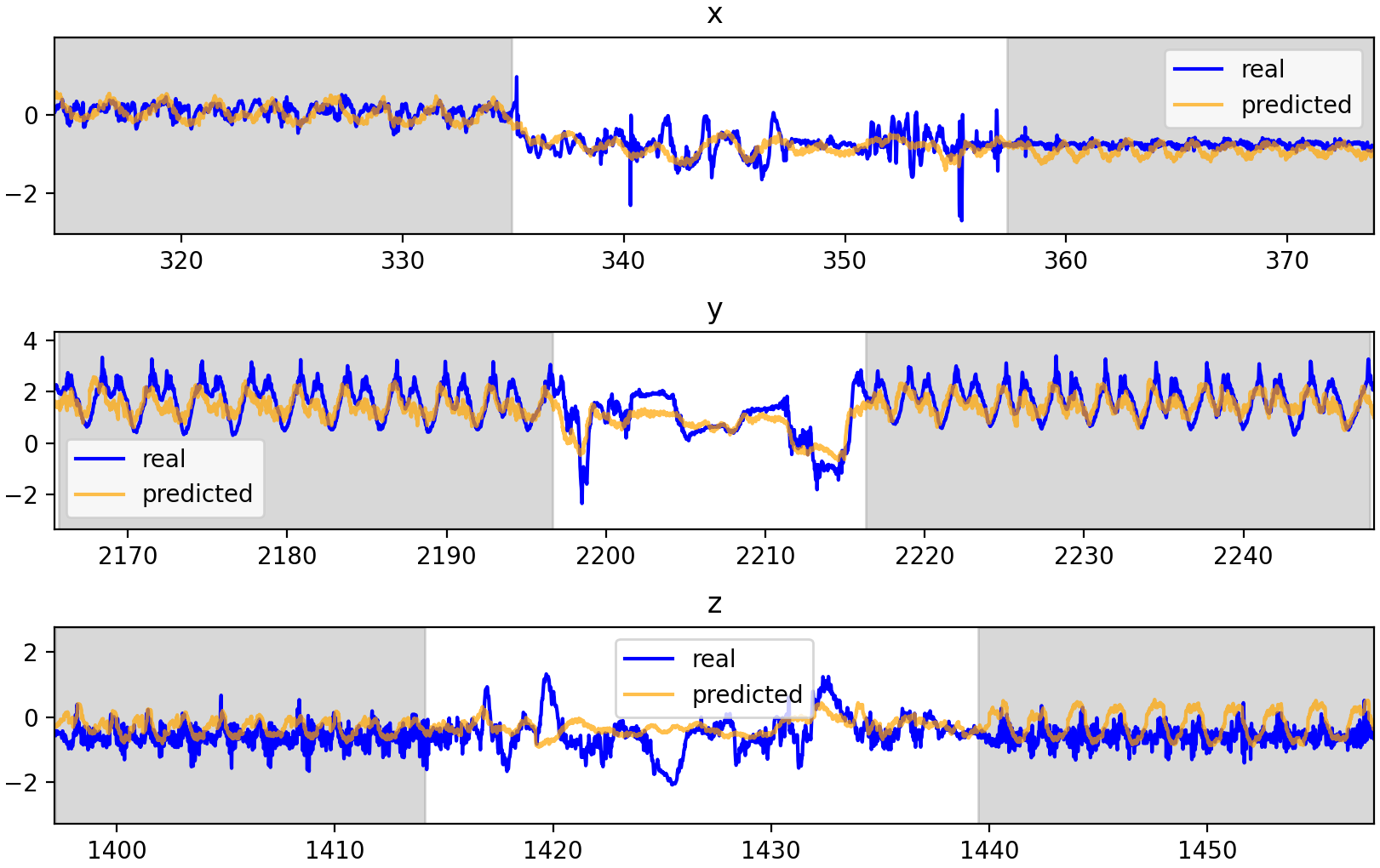}
  \hspace*{\fill} 
  \includegraphics[width=0.49\textwidth]{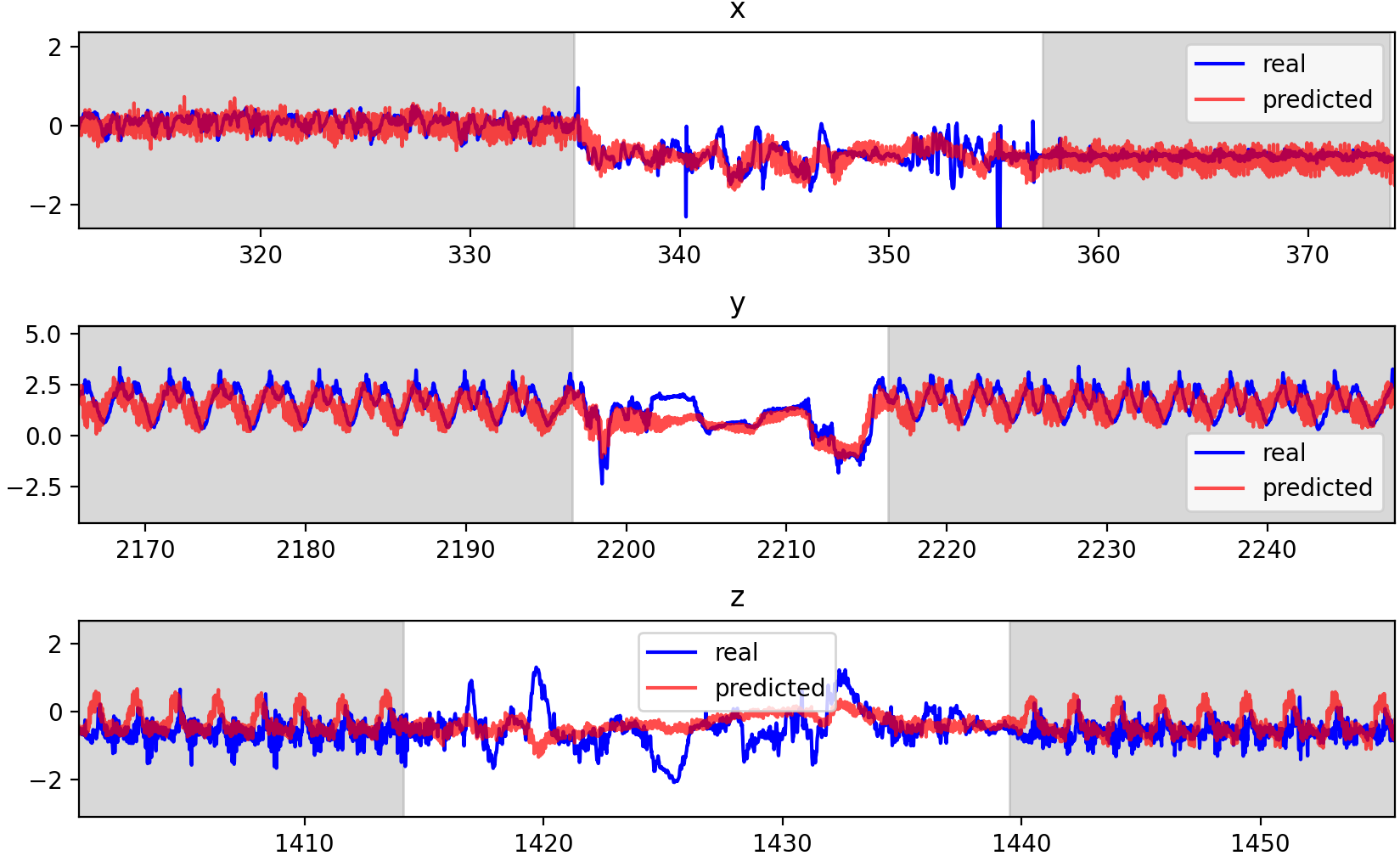}
  \caption{Qualitative Comparison of Regression Models. This figure contrasts the performance of a regression model trained within the end-to-end pipeline (right) against one trained independently (left). The real IMU accelerometer data is represented in blue, while the predictions from the regression model trained in the end-to-end pipeline are depicted in red. Predictions from the independently trained regression model are shown in orange. Both visualizations are based on identical time windows for a direct comparison.}
  \label{fig:side-by-side}
\end{figure*}

\subsection{Results}
\begin{table}[!t]
  \centering
  \caption{Comparison of Method Performance for MM-fit}
  \label{tab:method_comparison_mmfit}
  \begin{tabular}{lcc}
    \toprule
    Method & F1 Score & Accuracy \\
    \midrule
    IMUTube \cite{kwon2020imutube}  &  $0.7697  \pm 0.0019$
   &   $0.8981 \pm 0.0019$
    \\
    \midrule
    Baseline (real data) & $0.9025 \pm 0.0323$ & $0.9559 \pm 0.0125$ \\
    Regression-first & $0.8848 \pm 0.0183$ & $0.9490 \pm 0.0070$ \\
    Proposed Method ($\beta = 10$) & $0.9131 \pm 0.0266$ & $0.9494 \pm 0.0065$ \\
    \textbf{Proposed Method ($\beta = 0$)} & \textbf{$0.9196 \pm 0.0038$} & \textbf{$0.9618 \pm 0.0017$} \\
    \bottomrule
  \end{tabular}
\end{table}

\begin{table}[!t]
  \centering
  \caption{Comparison of Method Performance on the UTD-MHAD \cite{chen2015utd}. Results marked with * are reported performance from the reference papers.}
  \label{tab:method_comparison_mhad}
  \begin{tabular}{lcc}
    \toprule
    Method & F1 Score & Accuracy \\
    \midrule
    Chen et al. \cite{chen2015utd} * & - & 0.6720 \\
    Memmesheimer et al. \cite{memmesheimer2020gimme} *& - & 0.7286 \\
    \midrule
    Baseline (real data) & $0.6285 \pm 0.0171$ & $0.6702 \pm 0.0137$ \\
    Regression-first & $0.6650 \pm 0.0204$ & $0.6911 \pm 0.0121$ \\
    Proposed Method ($\beta = 10$) & $0.7342 \pm 0.0131$ & $0.7581 \pm 0.0020$ \\
    \textbf{Proposed Method ($\beta = 0$)} & \textbf{0.7388} $\pm$ \textbf{0.0101} & \textbf{0.7635} $\pm$ \textbf{0.0081} \\
    \bottomrule
  \end{tabular}
\end{table}

This section delves into the comparative analysis of the proposed method against the baselines. The discussion centers around the implications of the findings in terms of F1 score and accuracy, providing insights into the efficacy of synthetic data in training models for Human Activity Recognition (HAR).

\textbf{Proposed and Baseline Methods Comparison:}
As we can see in \cref{tab:method_comparison_mmfit} and \cref{tab:method_comparison_mhad}, our proposed approach provides improvements in F1 scores and accuracy metrics, which strongly suggests that the integration of synthetic accelerometer data within the training process substantially enhances the performance of both the feature extraction and classification models. This observation is particularly pronounced in the context of our end-to-end pipeline, which seamlessly incorporates synthetic data alongside real sensor inputs. 


\begin{table}[!t]
  \centering
  \caption{Comparison of Sensor Generation Quality for MM-Fit}
  \label{tab:regression_comparison_MMFit}
  \begin{tabular}{lcc}
    \toprule
    Regression Training & MSE in the Test set \\
    \midrule
    Regression-first & $0.4890 \pm 0.0123$  \\
    \textbf{End-to-end pipeline} & \textbf{$0.4620 \pm 0.0055$} \\
    \bottomrule
  \end{tabular}
\end{table}

\begin{table}[!t]
  \centering
  \caption{Comparison of Sensor Generation Quality for UTD-MHAD \cite{chen2015utd}}
  \label{tab:regression_comparison_MHAD}
  \begin{tabular}{lcc}
    \toprule
    Regression Training & MSE in the Test set \\
    \midrule
    Regression-first & $0.3910 \pm 0.0134$  \\
    \textbf{End-to-end pipeline} & \textbf{$0.3281 \pm 0.0122$} \\
    \bottomrule
  \end{tabular}
\end{table}

\textbf{Proposed and Regression-first Methods Comparison:}
In addition to juxtaposing the proposed method with the baseline, we explored an alternative scenario wherein the regression model is first trained independently to generate synthetic accelerometer data, which is then used alongside real data to train the feature extraction and classification models. This sequential approach, while theoretically sound, did not yield the same level of improvement as the integrated end-to-end pipeline. In fact, it is interesting to notice that performing the regression first did not improve results when compared to the baseline. In our tests we have access to the full training set and thus it is possible that the regression model alone could not generate simulated data of sufficient quality. The discrepancy (see again \cref{tab:method_comparison_mmfit,tab:method_comparison_mhad}) in performance can be attributed to the dynamic feedback loop established in the end-to-end training process, where the simultaneous adaptation of the regression model and the classification framework to each other's outputs fosters a more synergistic learning environment. This interdependence ensures that the synthetic data is not only accurate but also optimally aligned with the objectives of the feature extraction and classification tasks. 

This gap can be seen quantitatively when comparing the MSE of the generated data in the test set. As we can see in \cref{tab:regression_comparison_MMFit,tab:regression_comparison_MHAD}, our regression model provided a smaller mean MSE in the test set along with a smaller standard deviation. We can also see qualitatively in \cref{fig:side-by-side} that our approach provides better coverage of high-frequency components in the signal, even if this increases the overall noise in the signal. 

\textbf{Proposed and Existing Methods Comparison:}
We conducted a comparative analysis between our method and IMUTube \cite{kwon2020imutube} on the MM-Fit dataset. Given that our method generates sensor data using the original MM-Fit dataset exclusively, we followed the same protocol as IMUTube. Specifically, we utilized the pipeline outlined in \cite{kwon2020imutube} to create simulated accelerometer data for the left wrist using the original MM-Fit videos. This serves as a baseline measure of the quality of simulated data for this dataset, as IMUTube is not explicitly designed to optimize the quality of generated sensor data, unlike our regression model. Following the acquisition and calibration steps outlined in \cite{kwon2020imutube}, we trained our baseline model using a combination of half real sensor data and half IMUTube-generated data for each batch.

As depicted in \cref{tab:method_comparison_mmfit}, our proposed method outperforms IMUTube, which performs worse than the baseline. This is consistent with our previous findings: the quality of simulated data affects classification performance and, therefore, it is reasonable that simply applying IMUTube to the videos from the dataset itself only degrades the classifier performance. This can be overcome if one is using external videos to obtain additional data or, as is the case for our method, there is end-to-end optimization of sensor generation and activity recognition. 

In addition, we evaluated the proposed method against existing methods, Chen et al. \cite{chen2015utd} and Memmesheimer et al. \cite{memmesheimer2020gimme}, on the UTD-MHAD dataset. As shown in \cref{tab:method_comparison_mhad}, the results by the existing methods were reported performance from the references, and the proposed method outperformed the existing methods. 

\textbf{Implications and Future Directions:}
The findings from this study underscore the critical role of synthetic data in enhancing HAR systems, particularly in scenarios plagued by the scarcity of labeled datasets. The end-to-end pipeline proposed herein not only demonstrates the feasibility of such an approach but also sets a new benchmark for the integration of synthetic and real data in training sophisticated machine learning models. However, it is important to acknowledge that the accuracy of the generated sensor data is inherently tied to the quality of the input pose sequences. In this work, we utilized the pose sequences provided by the open-access benchmark dataset; however, in real-world settings, accurate 3D pose estimation techniques are crucial and can significantly impact the quality of the generated sensor data. Thus, in future work, we plan to explore more robust 3D pose estimation techniques and conduct sensitivity analyses to evaluate the robustness of the proposed method to variations in pose estimation accuracy.


\section{Conclusions}
\label{sec:conclusion}

In this study, we proposed an innovative approach to HAR by introducing a regression model, Pose2IMU, for accelerometer data and employing a novel combination of weighted loss functions. Our proposed method significantly enhanced the training and performance of activity classification models by incorporating synthetic accelerometer data derived from 3D skeleton poses. The use of TCN blocks tailored for processing 3D joint positions, along with adjustments in activation functions and optimization strategies, demonstrated a clear improvement in model accuracy and F1 scores compared to baseline methodologies.

The results affirmed the potential of synthetic data augmentation and sophisticated loss functions in overcoming the challenges posed by the limited availability of labeled HAR datasets. By effectively leveraging synthetic data, our method not only improved the depth and breadth of training data but also introduced a novel perspective on feature extraction and classification in the context of wearable sensor data.

Looking ahead, there are several avenues for extending this work. 
Exploring alternative end-to-end pipelines that incorporate multi-task learning could offer additional insights into the simultaneous optimization of related tasks, potentially leading to further improvements in HAR systems. In addition, we can adopt an adversarial learning scheme \cite{suh2022adversarial} between generated sensor data and real sensor data to not only match the generated sensor data with the real sensor data but also estimate the distributions of the real sensor data to improve the HAR performance.

\section*{Acknowledgments}
\label{sec:ack}
The research reported in this paper was supported by the BMBF in the project VidGenSense (01IW21003) and Carl-Zeiss Stiftung under the Sustainable Embedded AI project (P2021-02-009). 

\bibliographystyle{IEEEtran}
\bibliography{ref}

\end{document}